\documentclass[twocolumn,showpacs,preprintnumbers,amsmath,amssymb]{revtex4}
\usepackage{graphicx}
\usepackage{dcolumn}
\usepackage{color}
\usepackage{bm}
\usepackage{amsmath,amssymb,graphicx}
\usepackage{epsfig}
\usepackage{enumerate}
\usepackage{array}
\usepackage{multirow}

\begin{document}

\title
{DNA breathing dynamics: Analytic results for distribution functions of relevant Brownian functionals}
\author{Malay Bandyopadhyay$^1$, Shamik Gupta$^2$ and Dvira Segal$^1$}
\affiliation{$^1$Chemical Physics Theory Group, University of Toronto, 80, Saint George Street, Ontario M5S 3H6, Canada\\
$^2$Department of Physics of Complex Systems, Weizmann Institute of Science, Rehovot 76100, Israel}

\vskip-2.8cm
\date{\today}
\vskip-0.9cm

\begin{abstract}
We investigate DNA breathing dynamics by suggesting and examining several
different Brownian functionals
associated with bubble lifetime and reactivity.
Bubble dynamics is described as an overdamped random walk in the number of
broken base pairs. The walk
takes place on the Poland-Scheraga free energy landscape.
We suggest several probability distribution functions that characterize 
the breathing process, and adopt the recently studied backward Fokker-Planck method 
and the path decomposition method as elegant and flexible tools for deriving
these distributions.
In particular, for a bubble of an initial size $x_0$, we derive analytical expressions for (i) 
the distribution $P(t_f|x_0)$ of the first-passage time $t_f$, characterizing the bubble lifetime, 
(ii) the distribution $P(A|x_0)$ of the area $A$ till the first-passage time, providing 
information about the effective reactivity of the bubble to processes within the DNA, 
(iii) the distribution $P(M)$ of the maximum bubble size $M$ attained before the first-passage time, 
and (iv) the joint probability distribution $P(M,t_m)$ of the maximum bubble size $M$ and the time $t_m$ 
of its occurrence before the first-passage time. 
These distributions are analyzed in the limit of small and large bubble sizes. 
We supplement our analytical predictions with direct numerical simulations of the related Langevin equation, 
and obtain a very good agreement in the appropriate limits. 
The nontrivial scaling behavior of the various quantities analyzed here can, 
in principle, be explored experimentally.
\end{abstract}

\pacs{87.14.gk, 87.10.Mn, 02.50.-r, 05.40.-a}

\maketitle

\section{Introduction}
\label{intro}

The Watson-Crick double helix structure of DNA derives its stability from the 
phosphodiester bonds 
in the single-stranded sugar backbone, and from the hydrogen bonds between complementary base pairs on 
opposite strands \cite{DNA1,DNA2}. In practice, access to the inside of the double helix, and therefore, 
the unzipping of a specific region of base pairs is essential for all 
physiological processes involving DNA, e.g., for replication, transcription, and protein binding \cite{DNA3}.

Several mechanisms, like heating \cite{DNA-heat}, changing the pH of the environment \cite{Poland}, and
application of external force \cite{DNA-force} can lead to unzipping of the
double-stranded DNA. This phenomenon is referred to as DNA denaturation. 
The process occurs progressively, starting with the double strand separating locally into
single strands to form loops, or, ``bubbles''. These bubbles fluctuate in size through stepwise zipping and unzipping of the base pairs at the two zipper forks where the bubble connects to the double strand. At low
temperatures, bubbles once formed eventually close again in time. With the
increase of temperature, however, the bubbles grow in size in time to ultimately coalesce with neighboring bubbles 
and complete the denaturation process. The melting temperature $T_m$ is
defined as that at which half of the DNA molecule is denatured,
and has typical values $\sim 70 - 100^{\circ}$C for standard salt solutions \cite{Poland}. 
The number of bubbles varies from only a few ones well below $T_m$ up to several hundreds close to 
$T_m$. 

Breathing dynamics, referring to the dynamics of fluctuating DNA bubbles, has been a topic of intense research for many years \cite{DNA3, DNA-heat}. It has recently
regained interest with the development of new experimental tools that allow for the direct observation of the dynamics of a single DNA molecule \cite{Oleg-rev,Oleg}.
On the theoretical side, various methods have been used to study different aspects of the 
breathing process \cite{Special, Metzler-rev}, and to investigate the interaction of the DNA 
with binding proteins: the master equation approach \cite{master1,master2}, 
a stochastic Gillespie scheme \cite{Banik}, 
the Fokker-Planck equation approach based on the Poland-Scheraga free energy function \cite{Poland,FP1,FP2,FP3},
and  stochastic dynamic simulations based on the Dauxios-Peyrard-Bishop model \cite{Bishop,FP4}. 
Specifically, the thermally-induced denaturation problem has been recently studied by mapping 
it onto a quantum Coulomb problem \cite{Metzler-col,Wu}.
These studies have enhanced our understanding of general aspects of both polymer dynamics 
as well as specific biochemical processes. For recent advances, see \cite{Special}.

DNA breathing occurs on a timescale shorter than the equilibration time of the single strands 
forming the bubbles \cite{Oleg}. Based on this observation, breathing dynamics may be regarded as a random walk 
in the one-dimensional coordinate $x$, the number of broken base pairs. 
Ignoring heterogeneity in the DNA structure,
this random walk may be modeled as a noisy overdamped motion at a finite temperature $T$ 
on the Poland-Scheraga free energy landscape, 
${\mathcal F}(x) \sim \gamma x+c k_BT \ln x$ \cite{Metzler-rev, Metzler-col}. 
The parameters $c>0$ and $\gamma$ (which can be of either sign) are defined later in the paper. 
As we show below, this form of the free energy implies a crossover scale 
$x_\mathrm{ch}$. For small bubbles ($x < x_\mathrm{ch}$), 
the random walk takes place in a potential $\sim \ln x$. In the opposite limit, 
the potential grows linearly with $x$, thereby implying a different dynamics. Furthermore, 
the sign of $\gamma$, as given by the system temperature, determines the nature 
(attractive/repulsive) of the potential. For $\gamma >0$, which happens at temperatures $T<T_m$, the potential is attractive for all bubble sizes, thereby implying an eventual bubble closure ($x=0$). On the other hand, above $T_m$, when $\gamma < 0$, a large bubble ($x > x_\mathrm{ch}$) evolves under a repulsive linear potential to grow in size toward full denaturation, while a small bubble ($x < x_\mathrm{ch}$) may still shrink in size to closure under the influence of the attractive $\ln x$ potential.

In this work, following the above picture, we complement previous single-bubble
studies by suggesting and analyzing new measures for exploring the DNA breathing process. 
We focus on several first-passage  ``Brownian" functionals \cite{commfunc}
of the fluctuating bubble, which eventually closes again,
and derive their  probability distribution functions (pdfs). We separately study
the small and large bubble limits, which exhibit different behaviors. The
functionals of interest characterize the lifetime of the bubble, 
the time-integrated bubble size till the first-passage time (the bubble ``area"),
its maximum size before closure, and the characteristic time for attaining the maximum size.
These measures are relevant for estimating the effective reactivity of the bubble, e.g., its efficiency for binding processes.

Another objective of this work is to advocate the use of the recently studied backward Fokker-Planck (BFP) method
\cite{Majumdar-rev} and the path decomposition (PD) method \cite{Majumdar-rev2},
which builds on the Feynman-Kac formalism \cite{Kac}, for exploring DNA bubble dynamics.
These techniques have been extremely useful in studying many aspects of 
classical Brownian motion, as well as for exploring related problems in computer science and astronomy \cite{majumdar1, majumdar2, Majumdar-rev}.
Here, for the first time, we adopt these elegant methods in the context of DNA breathing dynamics.
Using the BFP method, we derive and solve differential equations for the Laplace transforms of various Brownian functionals.
This is in contrast to the standard Fokker-Planck treatment, which yields the distribution function to obtain a bubble 
of a given size at a given time \cite{FP1,FP2,FP3,FP4}.
Utilizing the PD approach, we can calculate the distribution functions of interest
by splitting a representative path of the dynamics into parts, and then considering the weight of each part separately.
This is justified by the Markovian property of the dynamics.

In order to gain a qualitative understanding of the DNA breathing process,
we separately consider the cases of small and large bubbles. 
Our analysis reveals new scaling laws for the pdfs of various Brownian
functionals, which are evidently distinct for small and large bubbles.
We further compare our analytical predictions with direct numerical simulations of the corresponding Langevin 
equation and observe a very good agreement in the appropriate limits.

The paper is organized as follows. 
In Section \ref{section2}, we recall the random walk model and discuss the distribution functions 
of interest and their relevance to the DNA dynamics.
The BFP and the PD methods are also explained, along with a short description of the numerical technique adopted. 
In Section \ref{small}, we study the dynamics of small bubbles, and derive the probability 
distribution functions of several first-passage Brownian functionals. 
In Section \ref{large}, the dynamics of large bubbles is examined.
We draw our conclusions in Section \ref{conclusions}.


\section{Model, Quantities of interest, and Methods}
\label{section2}

\subsection{Model}
\label{model}
We follow the Poland-Scheraga approach,
and interpret bubbles as occurring due to free-energy changes to the double-helical ground state \cite{Poland}. 
Measuring the size of a bubble by the number of broken base pairs, and denoting this number by the 
continuous variable $x \ge 0$, the Poland-Scheraga free energy is given by \cite{Poland}
\begin{equation}
{\cal{F}}(x)=\gamma_0 +\gamma x+ c k_B T \ln x,
\label{eq:free}
\end{equation}
where $\gamma_0$ is the free energy barrier to form the initial bubble, while 
the term $\gamma x$ stands for the free energy required in breaking $x$ base pairs. 
The entropy loss in forming a closed polymer loop is taken into account by the term $ck_BT\ln x$, 
where $k_B$ is the Boltzmann constant, $T$ is the temperature, while $c$ is a universal constant determined by 
the loop configurations \cite{c-refs}. In Eq. (\ref{eq:free}), a cutoff at $x \sim 1$ is implied. 
The parameter $\gamma$ is assumed to have the simple form, $\gamma=\gamma_1(T_m- T)/T_m$, where $\gamma_1=4k_BT_r$, with $T_r=310 {\rm K}$ being the reference temperature. 

At finite temperatures, the stochastic dynamics of DNA breathing can be
modeled by the overdamped Langevin equation
\cite{Metzler-rev,Metzler-col},
\begin{equation}
\frac{dx}{d\tau}=-D\frac{d{\cal{F}}}{dx} + \xi(\tau).
\label{eq:lang1}
\end{equation}
Here, $\xi(\tau)$ is a Gaussian white noise with $\langle {\xi}(\tau)
\rangle=0$, and $ 
\langle\xi(\tau)\xi(\tau^{\prime})\rangle = 2Dk_BT\delta(\tau-\tau^{\prime})$. 
The kinetic coefficient $D$ has the dimension of $(k_BT)^{-1}s^{-1}$. 
Using the free energy (\ref{eq:free}) in Eq. (\ref{eq:lang1}), and redefining the time variable,  $t \equiv 2Dk_BT\tau$,
results in the equation
\begin{equation}
\frac{dx}{dt}=C_2-\frac{C_1}{x}+ \tilde{\xi}(t),
\label{eq:lang2}
\end{equation}
where  $C_1=c/2$, $C_2=\gamma_1(T-T_m)/(2k_BTT_m)$, and
\begin{eqnarray}
&&\langle \tilde{\xi}(t) \rangle=0, 
\nonumber\\
&&\langle\tilde{\xi}(t)\tilde{\xi}(t^{\prime})\rangle = \delta(t-t^{\prime}).
\label{eq:corr}
\end{eqnarray}
Equation (\ref{eq:free}) implies a  crossover scale,
\begin{equation}
x_\mathrm{ch}=\frac{C_1}{|C_2|},
\label{eq:x1}
\end{equation}
such that for small bubbles, $x < x_\mathrm{ch}$, the free energy is dominated by the entropic term $\sim \ln x$. Correspondingly, the Langevin dynamics (\ref{eq:lang2}) is essentially governed by the term $-C_1/x$. For large bubbles with $x > x_\mathrm{ch}$, the base-pair dissociation term $\sim \gamma x$ dominates the free energy, and correspondingly, it is the $C_2$ term which dictates the Langevin dynamics. For $T< T_m$, the Langevin dynamics occurs in an attractive potential for all bubble sizes, thereby ensuring eventual bubble closure. Above $T_m$, large bubbles with $x > x_\mathrm{ch}$ grow in size under a repulsive linear potential to ultimately yield full denaturation, while a small bubble with $x < x_\mathrm{ch}$ may evolve towards closure under the influence of the attractive $\ln x$ potential. We will utilize below the length scale (\ref{eq:x1}) in distinguishing between small and large bubbles. Note that at precisely the melting temperature $T_m$, when $C_2=0$ and the characteristic bubble size diverges, the Langevin dynamics becomes identical to that for small bubbles at all temperatures $T \ne T_m$. 
 
\begin{figure}[htbp]
{\hbox{\epsfxsize=80mm\epsffile{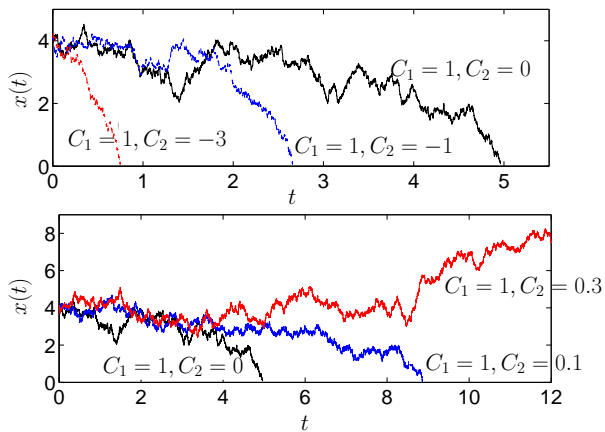}}}
\caption{(Color online) An illustration of several representative paths $x(t)$ following the 
time evolution of Eq. (\ref{eq:lang2}). 
All the paths begin at $x_0=4$. The values of the parameters $C_1$ and $C_2$
are marked in the figure. These paths have been generated by using a Brownian
simulation (see Section \ref{simulations}). }
\label{Fig0}
\end{figure}

Figure \ref{Fig0} depicts several characteristic paths evolving under 
Eq. (\ref{eq:lang2}) by adopting different values of the parameter $C_2$ with a fixed $C_1$. 
If $|C_2|>C_1$ and $C_2<0$ (i.e., $T<T_m$), the bubble closes sufficiently fast in comparison to the case $C_2=0$ (top panel). 
In the opposite limit, taking positive values for $C_2$, 
one observes the melting process reflected in the divergence of the first-passage time (bottom panel).


\subsection{Quantities of interest}
\label{quantitiesofinterest}

Our primary focus is on several first-passage Brownian functionals of experimental relevance. 
We will consider the following quantities and explore their pdfs for small and large bubbles.

(i) {\it Bubble lifetime.} The first-passage time pdf $P(t_f|x_0)$ i.e., the pdf of the time of closure for bubbles of initial size $x_0$,
provides information about bubble lifetime. 
A related quantity is the survival probability $C(x_0,t)\equiv 1-\int_{0}^{t} P(x_0|t_f)dt_f$
which can be inferred from experiments by measuring fluorescence correlations of a tagged DNA \cite{Oleg-rev,Oleg}.

\begin{figure}[htbp]
\begin{center}
{\hbox{\epsfxsize=80mm \epsffile{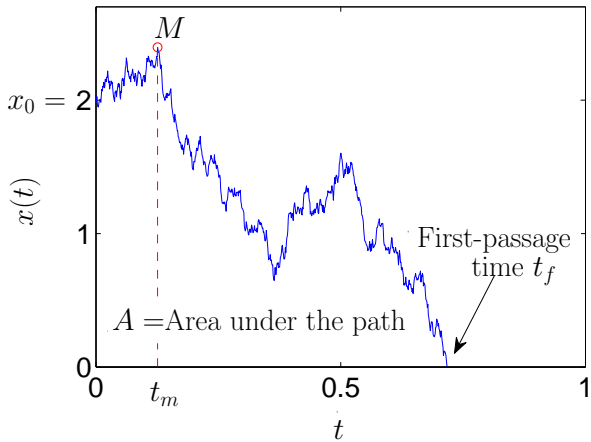}}} \caption{(Color online) 
An illustrative path $x(t)$ that begins at $x_0$ and evolves under 
Eq. (\ref{eq:lang2}). 
Here, $t_f$ marks the time the path crosses the origin for the first time (corresponding to bubble closure), 
$A$ is the area enclosed under the path,
$M$ and $t_m$ represent respectively the maximum value that the path reaches before the first-passage 
time, and the corresponding time of occurrence. 
This path has been generated by using a Brownian simulation (see Section \ref{simulations}) 
with $C_1=1$, $C_2=-1$, and $x_0=2$.} 
\label{Fig1}
\end{center}
\end{figure}

(ii) {\it Area under a path.} For the path described by Eq. (\ref{eq:lang2}), 
we define the area under the path before the first-passage time as $A=\int_0^{t_f}x(t')dt'$, 
see Fig. \ref{Fig1}, and calculate its pdf $P(A|x_0)$. 
This quantity is of interest since it provides a measure for the effectiveness 
of breathing-assisted processes, i.e., 
binding of proteins to the reactive sites of the DNA bases. 
As an example, consider a process that can take place only inside the double helix, on the single-stranded DNA. 
Let us assume that this process is facilitated with increasing bubble size, 
and that it requires a sufficiently long bubble lifetime. 
While the first-passage time distribution provides information about the average bubble lifetime,
it does not contain any hint of the average bubble size before closure. 
Similarly, $P(M)$, the distribution of the maximum bubble opening before closure
provides a measure for the bubble size, but it does not inform us about the
corresponding timescale.
Thus, we propose the pdf $P(A|x_0)$ of the area $A$ covered till the first-passage time 
as a useful quantity that provides a measure of bubble reactivity by containing
information about both size of the bubble and its characteristic lifetime.

(iii) {\it Maximum bubble size $M$.} 
Another proposed measure for quantifying bubble reactivity is the distribution of the maximum bubble size 
before the first-passage time, $P(M)$. 
Consider again a binding process taking place only inside the double helix.
Assume next that, due to geometrical constraints, the process may materialize only when the bubble is large enough.
If the timescale of this process is very short, shorter than the average bubble lifetime, 
a relevant measure for the bubble reactivity is its maximum opening before closure. 

Quantities (i), (ii) and (iii) will be calculated below by following the backward Fokker-Planck 
method discussed in Section \ref{BFPmethod}. 

(iv) {\it Maximum size $M$ and the corresponding time $t_m$.} 
The joint probability distribution function $P(M,t_m)$ will be investigated here by following the PD method, which builds on the Feynman-Kac formalism \cite{Majumdar-rev,Majumdar-rev2}; see Section \ref{PDmethod}. 
Using this pdf, one can further calculate the distribution function $P(t_m)$ of 
the time at which the bubble attains its maximum size before closure.
This latter pdf is of interest since it provides information about the (average) time of occurrence of the 
biggest bubble before closure. 
Processes taking place inside the DNA, facilitated by increased bubble size, will most likely occur around that time.

Figure \ref{Fig1} illustrates a typical path following Eq. (\ref{eq:lang2}). 
The path begins at $x_0$ and ends at 
the origin (bubble closure), staying positive in between. 
The various measures suggested above are also indicated in the figure.


\subsection{The backward Fokker-Planck (BFP) method}
\label{BFPmethod}

Following \cite{Majumdar-rev}, we recall here how to calculate the statistical properties of a 
Brownian functional, defined as
\begin{equation}
T=\int_0^{t_f} U(x(\tau))d\tau.
\label{eq:T}
\end{equation}
Here, $x(\tau)$ is a path representing the motion (\ref{eq:lang2}) which starts at $x_0$ at time $\tau=0$ 
and propagates up to  $\tau=t_f$, the first-passage time. In the above
equation, $U(x(\tau))$ is a specified function of the path, whose choice depends on the quantity of interest. 
For example, to compute the distribution of the first-passage time $t_f$, one chooses $U(x(\tau))=1$. 
For the area distribution till the first-passage time, one should consider $U(x(\tau))=x$. 
To find the pdf $P(T|x_0)$,  noting that the random variable $T$ can be only
positive for these choices of $U(x(\tau))$, 
one considers its Laplace transform
\begin{equation}
Q(x_0,p)=\!\int_0^{\infty}\!\!\!\!P(T|x_0)e^{-pT}dT=\langle e^{-p\int_0^{t_f} U(x(\tau))d\tau}\rangle,
\label{eq:Qdef}
\end{equation}
where the angular brackets denote averaging over all paths starting at $x_0$ at $\tau=0$ and ending at the 
first time they cross the origin. 
For simplicity of notation, in what follows, we suppress the variable $p$ in the function $Q(x_0,p)$. 
In order to derive a differential equation for $Q(x_0)$, we follow \cite{Majumdar-rev} 
and split the interval $\lbrack 0, t_f \rbrack$ into two parts. 
During the first interval $\lbrack 0,\Delta \tau \rbrack$,  
the path starts from $x_0$ and propagates up to $x_0+\Delta x$.
In the second interval $\lbrack \Delta \tau, t_f\rbrack$, the path starts at $x_0+\Delta x$ and reaches $0$ at $t_f$. Here, $\Delta \tau$ is a fixed, infinitesimally small time interval. 
We get, to leading order in $\Delta \tau$, 
$\int_0^{t_f}U(x(\tau))d\tau \approx U(x_0)\Delta \tau + \int_{\Delta \tau}^{t_f}U(x)d\tau$, and hence, from Eq. (\ref{eq:Qdef}), 
\begin{eqnarray}
Q(x_0)& \approx &e^{-pU(x_0)\Delta \tau}\langle Q(x_0+\Delta x)\rangle_{\Delta x} \nonumber\\
& \approx &(1-pU(x_0)\Delta \tau)\langle Q(x_0+\Delta x)\rangle_{\Delta x}.
\label{eq:Qint}
\end{eqnarray}
Now, the average denoted by the angular brackets is performed over all realizations of $\Delta x$. 
The dynamical equation (\ref{eq:lang2}) gives $\Delta x= F(x_0)\Delta \tau+{\tilde \xi(0)}\Delta \tau$, 
with $F(x_0)=C_2-C_1/x_0$. 
Substituting for $\Delta x$ in Eq. (\ref{eq:Qint}), 
expanding $Q(x_0+\Delta x)$ in powers of $\Delta \tau$, and averaging over the noise by using 
$\langle \tilde\xi(0) \rangle =0$ and $\langle \tilde\xi^2(0)\rangle=1/\Delta \tau$ for small $\Delta \tau$, 
one obtains, to lowest order in $\Delta \tau$, the ordinary differential equation,
\begin{equation}
\frac{1}{2}\frac{d^2Q(x_0)}{dx_0^2}+\left(C_2-\frac{C_1}{x_0}\right)\frac{dQ(x_0)}{dx_0}-pU(x_0)Q(x_0)=0.
\label{eq:Qgeneral}
\end{equation}
{\it Boundary conditions.} The above equation is valid for $x_0 \in \lbrack 0,\infty \rbrack$ 
with the following boundary conditions: 
(i) For an infinitesimally small bubble, $x_0\rightarrow 0$, the first passage time  vanishes, $t_f \rightarrow 0$, 
so that $Q(x_0=0)=1$. 
(ii) If the bubble is initially large, $x_0\rightarrow \infty$, the first passage time diverges, 
hence, $Q(x_0\rightarrow \infty)=0$.

We emphasize that the differential equation (\ref{eq:Qgeneral}), referred to as the backward Fokker-Planck equation 
\cite{Majumdar-rev}, directly provides us with the Laplace-transformed pdfs of
various quantities which are 
determined by the choice of  $U(x)$. 
In contrast, the standard Fokker-Planck method adopted in \cite{FP1,FP2,FP3,FP4} yields the
{\it density} distribution function $P(x,t)$ to obtain a loop of size $x$ at time $t$. 
Thus, these two approaches are distinct, providing complementary information.

\subsection{The path decomposition (PD) method}
\label{PDmethod}

The principle of this technique is simple: 
Since the motion in Eq. (\ref{eq:lang2}) is Markovian, a typical path can be split into, e.g., 
two parts. Then, the weight of the whole path is the  product of the weights of the two split parts \cite{Majumdar-rev2}.

The above idea allows us to calculate the joint probability distribution
$P(M,t_m)$ of the maximum bubble size $M$ 
and the time $t_m$ at which this maximum occurs before closure, given that the initial size of the bubble is 
fixed at $x_0\in [0,M]$. 
By integrating over $M$, one can further obtain the marginal distribution $P(t_m)$.
We compute $P(M,t_m)$ by splitting a typical path into two parts, 
before and after $t_m$, with the respective weights $W_L$ and $W_R$, so that
the weight $W$ of the whole path is 
\begin{eqnarray}
W=W_L \times W_R.
\end{eqnarray}
On the left side of $t_m$, the path propagates from $x_0$ at $t=0$ to $M-\epsilon$ at $t=t_m$, 
without ever attaining the value $0$ or $M$ during the interval $\lbrack 0,t_m \rbrack$ \cite{infi}. 
The weight $W_L$  can be determined by using a path integral treatment based on the Feynman-Kac formalism, 
as we explain below.
On the right side of $t_m$, the path starts from $M-\epsilon$ at $t=t_m$ and ends at the origin at $t_f$ 
(with $t_f\ge t_m$), without crossing either the level $M$ or the level $0$ in
between. At the end of the calculation, one needs to take the limit $\epsilon
\rightarrow 0$.

The calculation of $W_R$ will be explained in Sections  \ref{tmM-small} and \ref{tmM-large}.
We explain here in some detail the calculation of $W_L$.
Since the white noise in Eq. (\ref{eq:lang2}) is Gaussian, the probability of a path is given by
\begin{eqnarray}
P\lbrack\lbrace x(\tau)\rbrace\rbrack \propto \exp\left\lbrack-\frac{1}{2}\int_0^td\tau\Big(\frac{dx}{d\tau}+\frac{C_1}{x}-C_2\Big)^2\right\rbrack.
\end{eqnarray}
The weight $W_L$ is then given as a sum over contributions from all possible paths,
\begin{widetext}
\begin{eqnarray}
&&\!\!\!\!\!\!\!\!\!\!\!\!W_L \propto \int\limits_{x(0)=x_0}^{x(t_m)=M-\epsilon}{\cal{D}}x(\tau)\exp\left\lbrack-
\frac{1}{2}\int_0^{t_m} d\tau
\Big(\frac{dx}{d\tau}+\frac{C_1}{x}-C_2\Big)^2\right\rbrack\prod_{\tau=0}^{t_m} \theta[x(\tau)]\prod_{\tau=0}^{t_m} \theta[M-x(\tau)]
\nonumber\\
&&\!\!\!\!\!\!\!\!\!\!\!\!=\Big(\frac{x_0}{M-\epsilon}\Big)^{C_1}e^{C_2(M-\epsilon-x_0)}\!\!\!\!\!\!\int\limits_{x(0)=x_0}^{x(t_m)=M-\epsilon}\!\!\!\!\!\!{\cal{D}}x(\tau)\exp\left[-
\int_0^{t_m} d\tau \left\{\frac{1}{2}\Big(\frac{dx}{d\tau}\Big)^2+\frac{1}{2}\Big(\frac{C_1}{x}-C_2\Big)^2\right\}\right]\prod_{\tau=0}^{t_m} \theta[x(\tau)]\prod_{\tau=0}^{t_m} \theta[M-x(\tau)]. \nonumber \\ 
\label{eq:Kac}
\end{eqnarray}
\end{widetext}
In the above equations, the terms $\prod_{\tau=0}^{t_m} \theta[x(\tau)]$ and $\prod_{\tau=0}^{t_m} \theta[M-x(\tau)]$ 
enforce the requirements that the path does not cross either the level $0$ or the level $M$ for times between $0$ and $t_m$. 
Following Feynman-Kac \cite{Kac}, the path integral in Eq. (\ref{eq:Kac}) 
is identified with the propagator $\langle M-\epsilon|e^{-\hat{H}t_m}|x_0\rangle$, 
corresponding to the quantum Hamiltonian $\hat{H}$ of a single particle of unit mass, 
\begin{equation}
\hat{H}=-\frac{1}{2}\frac{d^2}{dx^2}+V(x),
\label{eq:H}
\end{equation}
with $\hbar=1$. The  potential energy $V(x)$ is given by
\begin{equation}
V(x)=\left\{ \begin{array}{ll}
          \frac{1}{2}\Big(\frac{C_1}{x}-C_2\Big)^2 & \mbox{if~} 0 < x < M, \\
          \infty & \mbox{if~} x=0  \mbox { or }  x=M. 
          \end{array}
          \right.
\label{eq:V}
\end{equation}
Note that the infinite potential energy at $x=0$ and at $x=M$ enforces the requirement that the path never crosses 
either the level $0$ or the level $M$. Finally, we get
\begin{eqnarray}
&&\!\!\!\!\!\!\!\!\!\!\!\!W_L \propto \Big(\frac{x_0}{M-\epsilon}\Big)^{C_1}e^{C_2(M-\epsilon-x_0)}\langle M-\epsilon|e^{-\hat{H}t_m}|x_0\rangle \nonumber \\
&&\!\!\!\!\!\!\!\!\!\!\!\!=\Big(\frac{x_0}{M-\epsilon}\Big)^{C_1}e^{C_2(M-\epsilon-x_0)}\sum_{p=1}^\infty e^{-E_pt_m}\psi_p(M-\epsilon)\psi_p(x_0), \nonumber \\
\label{eq:WL}
\end{eqnarray}
where $\psi_p(x)$ and $E_p$ are the eigenfunctions and eigenenergies, respectively, of the Hamiltonian 
$\hat{H}$ in Eq. (\ref{eq:H}). As a result of the infinite potential barrier at
$x=0$ and at $x=M$, the eigenfunctions satisfy $\psi_p(x=0,M)=0$. 

\subsection{Simulations}
\label{simulations}

The statistical properties of  Brownian functionals studied here can be numerically obtained 
by integrating the overdamped Langevin equation (\ref{eq:lang2}).  
Using a second-order stochastic Runge-Kutta algorithm \cite{algo}, we update
the trajectory by following the rule,
\begin{eqnarray}
x(\Delta t)&=&x_0+\frac{1}{2}\left[ F(x_0)+F(x_0+F(x_0) \Delta t +\Gamma_0)\right]\Delta t
\nonumber \\
&&+\Gamma_0,  
\end{eqnarray}
where $F(x)=C_2-C_1/x$. 
Here, $\Gamma_0$ is a random number sampled from a Gaussian distribution with zero mean and width 
given by  $\langle \Gamma_0^2\rangle = \Delta t$. 
For all simulations presented in this work, we take $\Delta t=10^{-3}$,  unless stated otherwise. 
We generate a large set of paths, all starting at a particular $x_0$ and ending close to the origin 
(within a preassigned numerical tolerance value). 
Averaging over an ensemble, we generate various pdfs which we compare with our analytical results.


\section{Small Bubble Dynamics}
\label{small}

We begin our analysis by considering small bubbles, $x < x_\mathrm{ch}$, at all temperatures, $T \ne T_m$. The analysis is also valid for bubbles of all sizes at precisely the melting temperature $T_m$. 
In these cases, the nonlinear entropic term in the free energy (\ref{eq:free}) dictates the dynamics, resulting in the Langevin equation,
\begin{equation}
\frac{dx}{dt}=-\frac{C_1}{x}+\tilde{\xi},
\label{eq:langsmall}
\end{equation}
where $\langle\tilde{\xi}(t)\rangle=0$ and 
$\langle \tilde{\xi}(t)\tilde{\xi}(t^{\prime}) \rangle=\delta(t-t^{\prime})$. 
The pdfs $P(t_f|x_0)$ and $P(A|x_0)$ are obtained by the BFP method, where
the differential equation that needs to be solved is given by Eq. (\ref{eq:Qgeneral}) with $C_2=0$,
\begin{equation}
\frac{1}{2}\frac{d^2Q(x_0)}{dx_0^2}-\frac{C_1}{x_0}\frac{dQ(x_0)}{dx_0}-pU(x_0)Q(x_0)=0.
\label{eq:Qsmall}
\end{equation}
The boundary conditions are (i) $Q(x_0\rightarrow\infty)=0$, and (ii) $Q(x_0 \rightarrow 0)=1$. 
We also derive analytical results for  $P(M)$ and $P(M,t_m)$, as explained below.

\subsection{First-passage time distribution: $P(t_f|x_0)$}
\label{tf-small}

We compute the distribution of $t_f$, the time at which the bubble closes for the first time, 
assuming its initial size is fixed at $x_0$, by substituting $U(x_0)=1$ in Eq. (\ref{eq:Qsmall}),
\begin{equation}
\frac{1}{2}\frac{d^2Q(x_0)}{dx_0^2}-\frac{C_1}{x_0}\frac{dQ(x_0)}{dx_0}-pQ(x_0)=0.
\label{eq:first-small}
\end{equation}
The general solution of Eq. (\ref{eq:first-small}) is \cite{gradshteyn}
\begin{equation}
Q(x_0)=x_0^\alpha\Big\lbrack A I_\alpha\Big(\sqrt{2p}x_0\Big)+BK_\alpha\Big(\sqrt{2p}x_0\Big)\Big\rbrack.
\label{eq:first-small-sol}
\end{equation}
Here, $I_\alpha(x)$ and $K_\alpha(x)$ are the modified Bessel functions of the
first and second kind, respectively. Also, 
$\alpha=C_1+1/2$, and $A$ and $B$ are arbitrary constants to be determined from the boundary conditions. 
Since for large $x$, $I_\alpha(x) \sim e^x/\sqrt{2\pi x}$ and $K_\alpha(x)\sim \sqrt{\pi/2x}~e^{-x}$ \cite{arfken}, 
in order to satisfy the condition $Q(x_0 \rightarrow \infty)=0$, we must have $A=0$. 
To satisfy the condition $Q(x_0 \rightarrow 0)=1$, we note that as $x \rightarrow 0$, $K_\alpha(x) \approx \Gamma(\alpha)2^{\alpha-1}/x^\alpha$ 
for $\alpha >0$ \cite{arfken}, which gives $B=(\sqrt{2p})^{C_1+1/2}/[\Gamma(C_1+1/2)2^{C_1-1/2}]$. 
Following these considerations, we get the particular solution,
\begin{equation}
Q(x_0)=x_0^{C_1+1/2}\frac{(\sqrt{2p})^{C_1+1/2}}{\Gamma(C_1+1/2)2^{C_1-1/2}}K_{C_1+1/2}\Big(\sqrt{2p}x_0\Big). 
\end{equation}
On taking inverse Laplace transform, we get \cite{bateman}
\begin{equation}
P(t_f|x_0)=\frac{x_0^{2C_1+1}}{\Gamma(C_1+1/2)2^{C_1+1/2}}t_f^{-C_1-3/2}e^{-x_0^2/2t_f},
\label{eq:Ptf1}
\end{equation}
as obtained earlier in \cite{bray, Metzler-col}.

Next, we compare the analytical prediction (\ref{eq:Ptf1}) with numerical simulations 
under the {\it full} bubble potential, including the $C_2$ contribution, 
in order to explore the regime of validity of the above result, see the top panel of Fig. \ref{Fig2}. 
The sampled trajectories all begin at $x_0=2$. 
When $x_\mathrm{ch}=C_1/|C_2| \gtrsim 10>x_0$; $C_2<0$, 
we observe a good agreement between numerics and analytical results. 
For $x_0 \sim x_\mathrm{ch}$, deviations occur since then the contribution of  $C_2$ 
cannot be neglected (inset). 

Besides the distribution (\ref{eq:Ptf1}), other related quantities of experimental relevance are the moments, 
$\langle t_f^k \rangle$, obtained from Eq. (\ref{eq:Ptf1}) as
\begin{eqnarray}
\langle t_f^k\rangle = \frac{x_0^{2k}}{2^k}\frac{\Gamma(C_1-k+1/2)}{\Gamma(C_1+1/2)} \text{~~for~~} k > C_1+1/2,
\end{eqnarray}
while $\langle t_f^k\rangle$ diverges for $k < C_1+1/2$. 
Another important quantity is  the persistence, or survival probability of the bubble, defined as
\begin{equation}
C(x_0,t)\equiv1-\int_0^t P(t_f|x_0)dt_f,
\label{eq:persist}
\end{equation}
where $\int_0^tP(t_f|x_0)dt_f$ sums up the probabilities of all events where the bubble closes in time $t$. 
This quantity can be resolved in experiments by measuring fluorescence correlations of a tagged bubble 
\cite{Oleg-rev, Oleg}. On plugging Eq. (\ref{eq:Ptf1}) into Eq.
(\ref{eq:persist}), we get 
\begin{eqnarray}
C(x_0,t)= 1- \frac{\Gamma(C_1+1/2, x_0^2/2t)}{\Gamma(C_1+1/2)},
\end{eqnarray}
where $\Gamma(s,x)=\int_x^\infty t^{s-1}\exp(-t)dt$ is the upper incomplete gamma function. 
This result agrees with that reported in \cite{Metzler-col}. It is easy to
derive the following asymptotic behaviors of $C(x_0,t)$:
In the limit $t\rightarrow \infty$, one has \cite{Metzler-col}
\begin{equation}
C(x_0,t) \approx
\frac{(x_0^2)^{C_1+1/2}}{(C_1+1/2)\Gamma(C_1+1/2)}t^{-C_1-1/2},
\end{equation}
while, in the limit $t\rightarrow 0$, one has
\begin{equation}
C(x_0,t) \approx 1-\frac{(x_0^2/2)^{C_1-1/2}}{\Gamma(C_1+1/2)}t^{1/2-C_1}
e^{-x_0^2/2t}.
\end{equation}


\subsection{Distribution of the area till the first-passage time: $P(A|x_0)$}
\label{A-small}

The area $A=\int_0^{t_f}x(t')dt'$ under the random motion (\ref{eq:langsmall})
can tell us about the readiness of the bubble to react. 
Here, the motion starts at $x_0$ and continues in time till the first-passage time.
Note that the quantity $A$ is not a geometric area, rather its units are  length $\times$ time.
To compute the related pdf, we substitute $U(x_0)=x_0$ in Eq. (\ref{eq:Qsmall}),
\begin{equation}
\frac{1}{2}\frac{d^2Q(x_0)}{dx_0^2}-
\frac{C_1}{x_0}\frac{dQ(x_0)}{dx_0}-px_0Q(x_0)=0,
\end{equation}
which has the general solution  \cite{bowman},
\begin{equation}
Q(x_0)=x_0^{C_1+1/2}\Big[A_1 J_{\nu}(iz)+A_2 J_{-\nu}(iz)\Big].
\end{equation}
Here, $z=(2/3)\sqrt{2p}~x_0^{3/2}$, 
$\nu=(2C_1+1)/3$, $J_\nu(x)$ is the Bessel function of the first kind, 
and $A_1$ and $A_2$ are arbitrary constants. 
Using $J_\nu(x)=i^\nu I_\nu(-ix)$, where $I_\nu(x)$ is the modified Bessel function of the first kind \cite{arfken},
gives 
\begin{equation}
Q(x_0)=x_0^{C_1+1/2}\Big[B_1 I_{\nu}(z)+B_2 I_{-\nu}(z)\Big],
\end{equation}
where $B_1$ and $B_2$ are arbitrary constants. 
Since for large $x$, $I_{\pm \nu}(x) \sim e^x/\sqrt{2\pi x}$ \cite{arfken},
to satisfy the boundary condition $Q(x_0\rightarrow\infty)=0$, 
we must have $B_1=-B_2$. To satisfy $Q(x_0 \rightarrow 0)=1$, we note that 
as $x_0 \rightarrow 0$, one has $I_\nu(x) \approx (x/2)^\nu/ \Gamma(\nu+1)$ \cite{arfken}, 
which yields $B_2=\Gamma(1-\nu)(\sqrt{2p}/3)^\nu$. We thus get
\begin{eqnarray}
Q(x_0)&=&\Big(\frac{z}{2}\Big)^\nu\Gamma(1-\nu)\Bigg[I_{-\nu}(z)-I_{\nu}(z)\Bigg] \nonumber \\
&=&\Big(\frac{z}{2}\Big)^\nu\frac{2}{\Gamma(\nu)}K_{\nu}(z), 
\label{eq:Qarea2}
\end{eqnarray}
where, in obtaining the last relation, 
we have used the identities, 
$K_\nu(x)=(\pi/2)[I_{-\nu}(x)-I_\nu(x)]/\sin (\nu\pi)$ 
and $\Gamma(\nu)\Gamma(1-\nu)=\pi/\sin (\nu \pi)$ \cite{arfken}. 
On taking inverse Laplace transform of (\ref{eq:Qarea2}), we obtain the
desired pdf \cite{bateman},
\begin{equation}
P(A|x_0)=\frac{2^{(2C_1+1)/3}x_0^{2C_1+1}}{3^{(4C_1+2)/3}\Gamma((2C_1+1)/3)}\frac{\exp(-2x_0^3/9A)}{A^{(2C_1+4)/3}}.
\label{eq:PA1}
\end{equation}
This expression nicely reproduces the numerical results obtained by simulating the 
Langevin equation (\ref{eq:lang2}) under the {\it full} potential with $x_0 < x_\mathrm{ch}$ and $C_2<0$, 
see the middle panel of Fig. \ref{Fig2}. 
The inset demonstrates an expected disagreement for larger bubbles with  $x_0 \sim x_\mathrm{ch}$. 

\subsection{Distribution of the maximum before the first-passage time: $P(M)$}
\label{M-small}

\begin{figure}[htbp]
{\hbox{\epsfxsize=90mm \epsffile{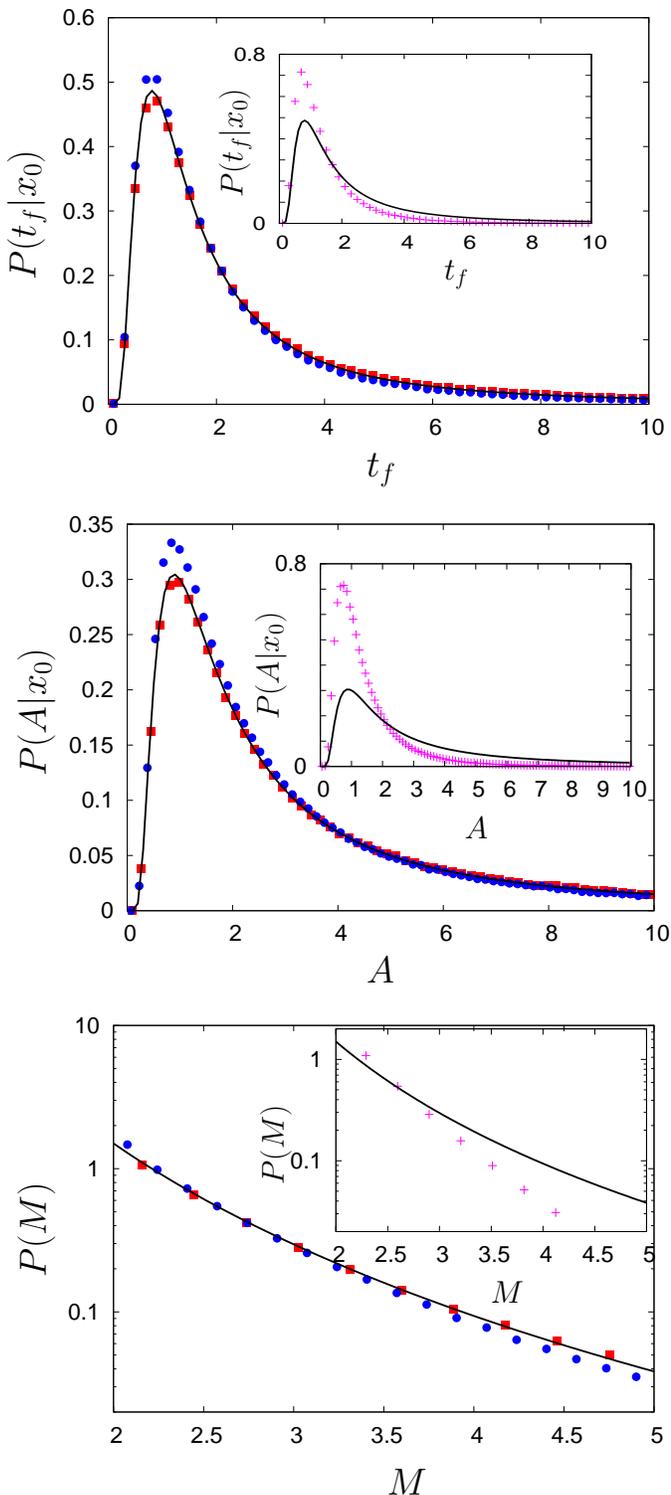}}}
\caption{(Color online) 
Numerical simulations below the melting temperature, $T<T_m$, 
for the pdf of the first-passage time (top), 
the pdf of the area till the first-passage time (center) 
and the pdf of the maximum size till the first-passage time (bottom). 
The parameters are $C_1=1$ and $C_2=0$ (red square), 
$C_1=1$ and $C_2=-0.1$ (blue dots). 
The initial bubble size is $x_0=2$ for all the cases. 
The analytic results in the small bubble approximation, (\ref{eq:Ptf1}), (\ref{eq:PA1}), 
(\ref{eq:PMsmall}) appear in black continuous lines. 
The three insets compare the analytic small bubble results (black continuous line) 
with numerical simulations using $C_1=1$ and $C_2=-0.5$ (purple $+$).
}
\label{Fig2}
\end{figure}

How large can the bubble grow before it closes, assuming an initial opening of $x_0$? 
This question is of interest in estimating the efficiency of processes that
can occur inside big loops only.
The relevant measure is provided by $P(M)$, the pdf of the maximum bubble size $M$ 
before its first closure, given that  $x_0 \in [0,M]$. We obtain this pdf by
following closely the procedure of \cite{satya-tf,Majumdar-rev2}. 
We first define a related function. 
Let $q(x_0)$ be the probability that the motion described by Eq. (\ref{eq:langsmall}) 
exits the interval $[0,M]$ for the first time through the origin. Thus, $q(x_0)$ is the cumulative probability that the maximum before the first-passage time is $\le M$. 
It is evident that this function satisfies two boundary conditions: 
(i) $q(0)=1$, and (ii) $q(M)=0$. 
Denoting by $\phi_{\Delta \tau}(\Delta x)$ the distribution function of a small displacement 
$\Delta x$ in time $\Delta \tau \rightarrow 0$, 
it follows from the Markovian property of the dynamics (\ref{eq:langsmall}) that
\begin{equation}
q(x_0)=\int q(x_0+\Delta x)\phi_{\Delta \tau}(\Delta x)d(\Delta x).
\end{equation}
On Taylor expanding $q(x_0+\Delta x)$ and averaging over $\Delta x=-(C_1/x_0)\Delta \tau+\tilde{\xi}(0)\Delta \tau$, using $\langle \tilde{\xi}(0) \rangle=0, \langle \tilde{\xi}^2(0) \rangle=1/\Delta \tau$, 
we get, to leading order in $\Delta \tau$, 
the equation $[(1/2)(d^2q(x_0)/dx_0^2)-(C_1/x_0)(dq(x_0)/dx_0)]\Delta \tau=0$. 
For arbitrary $\Delta \tau$, we obtain
\begin{equation}
\frac{1}{2}\frac{d^2q(x_0)}{dx_0^2}-\frac{C_1}{x_0}\frac{dq(x_0)}{dx_0}=0.
\end{equation}
Solving this equation with the above mentioned boundary conditions, we get
\begin{equation}
q(x_0)=1-\left(\frac{x_0}{M}\right)^{2C_1+1}.
\label{eq:qMsmall}
\end{equation}
The pdf of interest is obtained by differentiating $q(x_0)$ with respect to $M$, 
\begin{equation}
P(M)=\frac{(2C_1+1)x_0^{2C_1+1}}{M^{2C_1+2}};~~~~~~~~M \ge x_0.
\label{eq:PMsmall}
\end{equation}
In Fig. \ref{Fig2} (bottom), we compare this result with numerical simulations 
using the full potential.  We observe good agreement for $x_0 < x_\mathrm{ch}; C_2<0$, 
and an expected disagreement for $x_0 \sim x_\mathrm{ch}$ (inset).


\subsection{Joint pdf of the maximum $M$ and the corresponding time $t_m$ before the first-passage time: $P(M,t_m)$}
\label{tmM-small}

To compute the joint probability distribution of the maximum bubble size $M$ and the time $t_m$ at which the maximum occurs 
before closure, we adopt the PD method, see Section \ref{PDmethod} and \cite{Majumdar-rev2}. 
We split a path evolving under (\ref{eq:langsmall}) into two parts,
before  and after  the time $t_m$, with respective weights $W_L$ and $W_R$.
Due to the Markovian property, the weight $W$ of the whole path is given by the product of the weights for 
the two split parts, $W=W_L\times W_R$.
The weight $W_R$  can be obtained from Eq. (\ref{eq:qMsmall}). Recall that
$W_R$ is the weight of a path that starts at $M-\epsilon$ at time $t=t_m$ and
exits the interval $[0,M]$ for the first time through the origin. On the
other hand, $q(x_0)$ in Eq. (\ref{eq:qMsmall}) is the probability for a path starting at $x_0 \in [0,M]$ 
to exit the interval for the first time through the origin. We thus deduce
that $W_R=q(M-\epsilon)$, or, 
\begin{equation}
W_R=1- \frac{(M-\epsilon)^{(2C_1+1)}}{M^{2C_1+1}} = \frac{(2C_1+1)\epsilon}{M} + {\cal O}(\epsilon^2).
\label{eq:WRsmall}
\end{equation}
The second equality is derived by assuming $\epsilon$ to be infinitesimal. 
The weight $W_L$ is obtained from Eq. (\ref{eq:WL}) by substituting $C_2=0$,
\begin{equation}
W_L \propto \Big(\frac{x_0}{M-\epsilon}\Big)^{C_1}\sum_{p=1}^\infty e^{-E_pt_m}\psi_p(M-\epsilon)\psi_p(x_0).
\label{eq:WLsmall}
\end{equation}
Here, $\psi_p$ and $E_p$ are the solutions of the eigenequation,
\begin{equation}
\left[-\frac{1}{2}\frac{d^2}{dx^2}+\frac{C_1^2}{2x^2}\right]\psi=E\psi;
\,\,\,\,\,\,\, 0 < x < M,
\label{eq:diffeqsmall}
\end{equation}
subject to the condition $\psi(x=0,M)=0$. 
The general solution of this equation is
\begin{eqnarray}
\psi_p(x)=A\sqrt{x}J_\alpha\left(\sqrt{2E_p}x\right)+B\sqrt{x}Y_\alpha\left(\sqrt{2E_p}x\right), 
\end{eqnarray}
where $J_\alpha(x)$ and $Y_\alpha(x)$ are the Bessel functions of order $\alpha$ of the first and second kind, respectively, 
and $\alpha=\frac{1}{2}\sqrt{1+4C_1^2}$ \cite{gradshteyn}. 
Note that  $C_1$ is real, thus $\alpha>0$. 
Since for $x \to 0$, $Y_\alpha(x) \approx -(\Gamma(\alpha)/\pi)(2/x)^\alpha$ \cite{arfken}, 
we demand that $B=0$ for satisfying  $\psi(x=0)=0$. 
The other boundary condition results in the discrete eigenvalues $E_p$ such
that $\sqrt{2E_p}M=u_{\alpha p}$, where  
$u_{\alpha p}$ denotes the $p$-th zero of $J_\alpha(x)$.  
The constant $A$ is determined by requiring $\psi_p(x)$ to be normalized.
On using the identity, 
$\int_0^a d\rho \rho J_\alpha(u_{\alpha p}\rho/a)J_\alpha(u_{\alpha q}\rho/a)=\delta_{p,q}(a^2/2)\left[J_{\alpha+1}(u_{\alpha p})\right]^2$
\cite{arfken}, we finally get
\begin{equation}
\psi_p(x)=\frac{\sqrt{2x}}{M|J_{\alpha+1}(u_{\alpha p})|}J_\alpha\left(\frac{u_{\alpha p}x}{M}\right),
\label{eq:normeg}
\end{equation}
and the probability
\begin{eqnarray}
&&\!\!\!\!\!\!\!\!\!\!\!\!W_L \propto \Big(\frac{x_0}{M-\epsilon}\Big)^{C_1}\frac{2\sqrt{(M-\epsilon)x_0}}{M^2}\nonumber \\
&&\!\!\!\!\!\!\!\!\!\!\!\! \times \sum_{p=1}^\infty \frac{e^{-u_{\alpha p}^2t_m/(2M^2)}}{[J_{\alpha+1}(u_{\alpha p})]^2}J_\alpha\left(\frac{u_{\alpha p}(M-\epsilon)}{M}\right)J_\alpha\left(\frac{u_{\alpha p}x_0}{M}\right). \nonumber \\
\label{eq:genmatelement}
\end{eqnarray}
Next, we evaluate $W_L$ 
 to leading order in $\epsilon$ by Taylor expanding
 $J_\alpha\left(\frac{u_{\alpha p}(M-\epsilon)}{M}\right)$ and also using
 the result $J_\alpha '(u_{\alpha p})=-J_{\alpha+1}(u_{\alpha
p})$, which follows from the following 
identity: $J'_{\alpha}(z)=\frac{\alpha}{z}J_{\alpha}(z)-J_{\alpha+1}(z)$
\cite{gradshteyn}. We finally get
\begin{eqnarray}
W_L &\propto& \epsilon\frac{2x_0^{C_1+1/2}}{M^{C_1+5/2}}\sum_{p=1}^\infty u_{\alpha p} \frac{e^{-u_{\alpha p}^2t_m/(2M^2)}}{J_{\alpha+1}(u_{\alpha p})}J_\alpha\left(\frac{u_{\alpha p}x_0}{M}\right) 
\nonumber\\
&+& {\cal O}(\epsilon^2). 
\label{eq:WRsmallfinal}
\end{eqnarray}
The probability $P(M,t_m; \epsilon)$ of the whole path is the product of (\ref{eq:WRsmallfinal}) and (\ref{eq:WLsmall})
with a normalization constant $B(\epsilon)$, which is determined by requiring that
$\lim_{\epsilon \rightarrow 0} \int_0^\infty P(M,t_m; \epsilon)dt_m \rightarrow P(M)$, where $P(M)$ is given by Eq. (\ref{eq:PMsmall}),
\begin{eqnarray}
P(M,t_m; \epsilon)= B(\epsilon) W_L W_R.
\label{pmtmeps}
\end{eqnarray}
Using the identity $\sum_{p=1}^\infty J_\alpha(u_{\alpha p}x_0/M)/(u_{\alpha p}J_{\alpha+1}(u_{\alpha p}))
=x_0^\alpha/(2M^\alpha)$ for $0 \le x_0 < M$ \cite{prudnikov}, 
we get $B(\epsilon)=\frac{1}{2\epsilon^2}\left(\frac{x_0}{M}\right)^{C-C_1-1/2}$, where $C=2C_1+1-\sqrt{1+4C_1^2}/2$. 
Substituting for $B(\epsilon)$ in Eq. (\ref{pmtmeps}) and taking the limit
$\epsilon \rightarrow 0$, we get the desired probability, 
\begin{eqnarray}
&&P(M,t_m) = (2C_1+1) \frac{x_0^{C}}{M^{C+3}}\nonumber \\
&&\times \sum_{p=1}^\infty u_{\alpha p} \frac{e^{-u_{\alpha p}^2t_m/(2M^2)}}{J_{\alpha+1}(u_{\alpha p})}J_\alpha\left(\frac{u_{\alpha p}x_0}{M}\right).
\end{eqnarray}
For the free Brownian motion, taking $C_1=0$, this result reduces to that derived in \cite{Majumdar-rev2}. 


\section{Large bubble dynamics}
\label{large}

We study here the dynamics of large bubbles of size $x > x_\mathrm{ch}$, 
see Eq. (\ref{eq:x1}). 
In this limit, one can neglect the term $-C_1/x$ in the Langevin equation (\ref{eq:lang2}), and study the dynamics dictated by
\begin{equation}
\frac{dx}{dt}=C_2+\tilde{\xi}(t).
\label{eq:langlarge}
\end{equation}
This equation describes a one-dimensional random walk, $x(t)$, in the presence
of a constant drift, $C_2$. 
The probability distribution to find a bubble of size $x$ at time $t$ is a shifted Gaussian
$P(x,t) \sim e^{-(x-x_0-C_2t)^2/2t}$, where $x_0$ is the initial size of the bubble. 
Below the melting temperature ($C_2 <0$), the bubble shrinks and eventually disappears in time. 
Above $T_m$, when $C_2 >0$, the bubble grows in size as time passes. 
For large $C_2 >0$, the first-passage time diverges. 
Thus, our analysis below of various first-passage functionals is valid  for large bubbles at
temperatures below $T_m$ as well as above $T_m$, in the latter case while conditioned on a finite $t_f$.

For obtaining the pdfs, $P(t_f|x_0)$ and $P(A|x_0)$, we adopt the BFP method.
The relevant differential equation is obtained from Eq. (\ref{eq:Qgeneral}) by substituting $C_1=0$,
\begin{equation}
\frac{1}{2}\frac{d^2Q(x_0)}{dx_0^2}+C_2\frac{dQ(x_0)}{dx_0}-pU(x_0)Q(x_0)=0,
\label{eq:Qbig}
\end{equation}
 with the boundary conditions (i) $Q(x_0\rightarrow\infty)=0$, and (ii) $Q(x_0 \rightarrow 0)=1$. 
We also derive analytical results for $P(M)$ and $P(M,t_m)$.

\begin{figure}[htbp]
{\hbox{\epsfxsize=90mm \epsffile{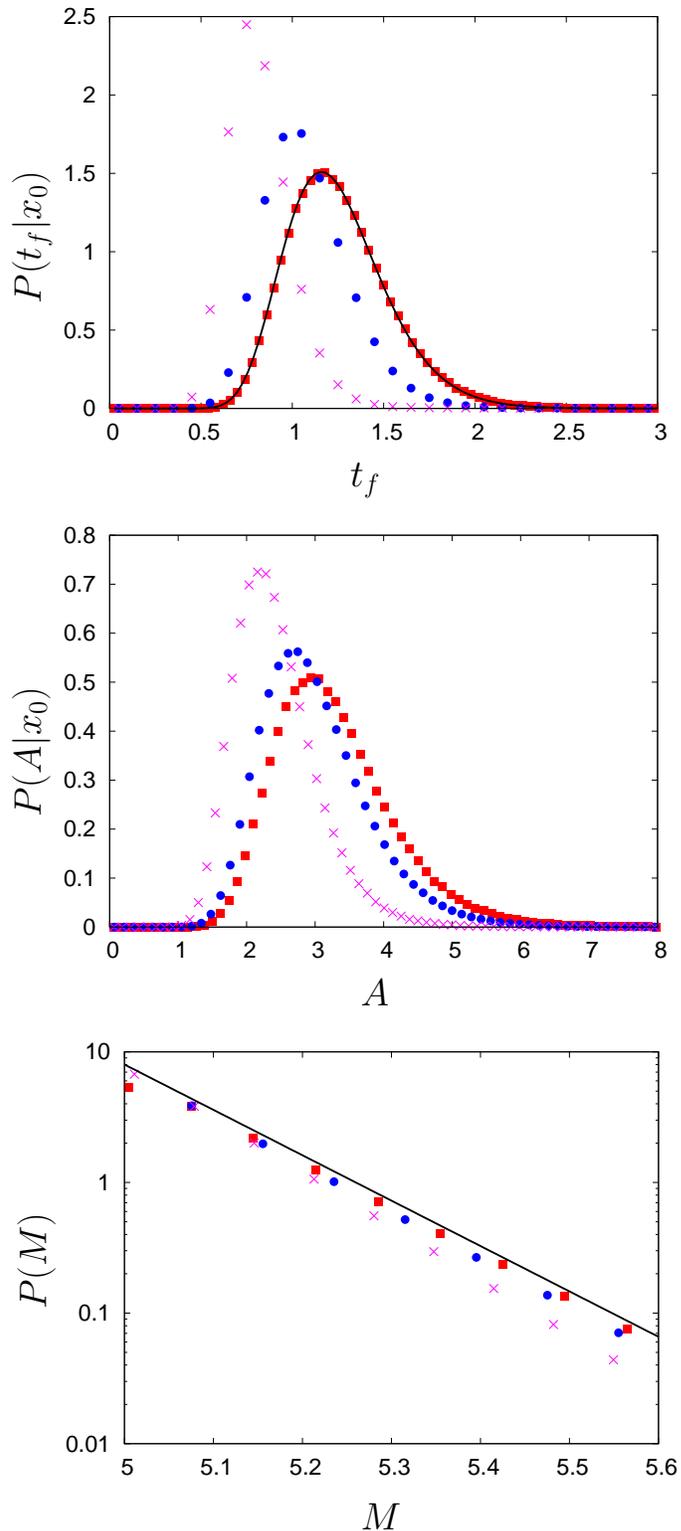}}}
\caption{(Color online) 
Results from numerical simulations below the melting temperature
for the pdf of the first-passage time (top), the pdf of the area till the first-passage time (center), and the pdf of the maximum size till the first-passage time pdf (bottom). 
The parameters are $C_1=0$ (red square), $C_1=1$ (blue dots), and $C_1=4$ (purple cross).
The initial bubble size is $x_0=5$ and  $C_2=-4$ in all the cases. 
The analytic results in the large bubble approximation appear in black continuous lines.
In the middle panel, we do not make a comparison with the analytic function for $P(A|x_0)$
since its explicit form is known only in the limit of small and large values of $A$.}
\label{Fig3}
\end{figure}

\begin{figure}[htbp]
{\hbox{\epsfxsize=85mm \epsffile{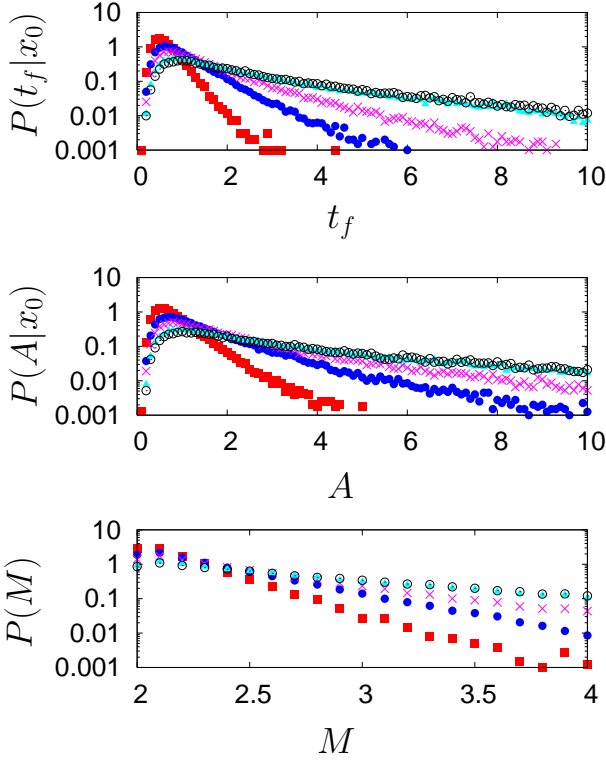}}}
\caption{(Color online)
Results from numerical simulations below and above $T_m$, showing the first-passage time pdf (top), 
the pdf of the area till the first-passage time (center),
and the pdf of the maximum till the first-passage time (bottom). 
The parameters are $C_2=-2$ (red filled square), $C_2=-1$ (blue filled circle),
$C_2=-0.5$ (purple cross), $C_2=0.1$ (light blue filled triangle), and $C_2=0.5$ (black circle). 
The initial bubble size is $x_0=2$ and $C_1=1$ in all cases.}
\label{Fig4}
\end{figure}

\subsection{First-passage time distribution: $P(t_f|x_0)$}
\label{tf-large}

This distribution has already been investigated in \cite{satya-tf} in a different context.
We thus omit the details of the calculation, but include the results for the
sake of completeness of our presentation.
The procedure involves solving  Eq. (\ref{eq:Qbig}) with $U(x_0)=1$ under the
boundary conditions,
then taking inverse Laplace transform of the solution to yield
\begin{equation}
P(t_f|x_0)=\frac{1}{\sqrt{2\pi}}\frac{x_0}{t_f^{3/2}}\exp\left[-\frac{(x_0+C_2t_f)^2}{2t_f}\right].
\label{eq:Ptf2}
\end{equation}
The moments $\langle t_f^k \rangle$ may be obtained by using the identity, $\int_0^\infty x^{\nu-1}e^{-\beta/x-\gamma x}dx=2(\beta/\gamma)^{\nu/2}K_\nu(2\sqrt{\beta \gamma})$ for Re$(\beta)>0$ and Re$(\gamma)>0$, where $K_\nu(x)$ is the modified Bessel function of the second kind \cite{gradshteyn}. One gets
\begin{equation}
\langle t_f^k\rangle = \sqrt{\frac{2}{\pi}}\Big(\frac{x_0}{|C_2|}\Big)^k(x_0|C_2|)^{1/2}e^{-C_2x_0}K_{k-1/2}(|C_2|x_0).
\end{equation}
Noting that $K_{1/2}(x)=\sqrt{\pi/(2x)}e^{-x}$, the mean first-passage time is given by 
$\langle t_f \rangle=x_0e^{-2C_2x_0}/C_2$ for $C_2 >0$, 
and by $\langle t_f \rangle=x_0/|C_2|$ for $C_2 < 0$. 
The survival probability, defined in Eq. (\ref{eq:persist}), is given by
\begin{equation}
C(x_0,t)=1-\frac{x_0}{\sqrt{2\pi}}e^{-C_2x_0}\int_0^t t_f^{-3/2}e^{-x_0^2/(2t_f)-C_2^2t_f/2}dt_f.
\end{equation}
For $C_2<0$ ($T < T_m$) and for large $t$, one has 
$C(x_0,t) \approx 1-(C_2x_0)/(2\sqrt{\pi})e^{-C_2x_0-x_0^2/(2t)}\Big[\Gamma(-1/2)-\Gamma(-1/2,C_2^2t/2)\Big]$. 
Using the result that $\Gamma(s,x) \rightarrow x^{s-1}e^{-x}$ as $x \rightarrow \infty$ \cite{arfken}, 
we get $C(x_0,t)\sim \sqrt{2/\pi}(x_0/C_2^2)t^{-3/2}\exp\left[-(x_0-|C_2|t)^2/(2t)\right]$. 

Figure \ref{Fig3} compares the analytic result for $P(t_f|x_0)$ with numerical simulations obtained by considering 
the full potential. Since, for $x_0>x_\mathrm{ch}$ with $C_2>0$, 
many trajectories have diverging $t_f$, we performed here numerical simulations only below the melting temperature,  
so that $C_2 <0$. 
In this case, with increasing $|C_2|$, the effective drift velocity towards bubble closure increases and 
bubbles quickly disappear in time.
The top panel demonstrates that the analytical prediction (\ref{eq:Ptf2}) agrees with simulation results 
for $x_0 > x_\mathrm{ch}; C_2 <0$.
Figure \ref{Fig4} further includes results from numerical simulations by considering the full bubble potential,
displaying the behaviors both below and above the melting temperature.
Upon increasing $C_2$ from negative values (i.e., for $T<T_m$) to positive values (i.e., for $T>T_m$), one notes that (i) the center of the pdf $P(t_f|x_0)$ is displaced to longer times, and that (ii) the bubble lifetime is significantly enhanced (see Fig. \ref{Fig4}, top panel). This corroborates with the physical picture that with increasing temperature, a bubble takes a longer time to disappear.

\subsection{Distribution of the area till the first-passage time: $P(A|x_0)$}
\label{A-large}

This quantity can be obtained by solving Eq. (\ref{eq:Qbig}) with
$U(x_0)=x_0$ with the appropriate 
boundary conditions, then deriving the inverse Laplace transform of the solution, see \cite{satya-tf,satya-A} for details. 
In particular, one obtains the following two limiting behaviors of the distribution $P(A|x_0)$:
For $A \rightarrow \infty$, one has 
\begin{eqnarray}
&&P(A|x_0)\approx \frac{e^{-C_2x_0}\sinh (|C_2|x_0)}{\sqrt{\pi}}\left(\frac{2}{3}\right)^{1/4}\left(\frac{|C_2|}{A}\right)^{3/4}\nonumber \\
&&\times \exp\left\{-\left(\frac{8}{3}\right)^{1/2}|C_2|^{3/2}A^{1/2}\right\}.
\label{eq:PA2a}
\end{eqnarray}
In the opposite $A \rightarrow 0$ limit, one gets \cite{satya-A}
\begin{equation}
P(A|x_0)\approx \frac{2^{1/3}}{3^{2/3}\Gamma(1/3)}\frac{x_0e^{-C_2x_0}}{A^{4/3}}e^{-2x_0^3/9A}.
\label{eq:PA2b}
\end{equation}
Note the distinct asymptotic forms in Eqs. (\ref{eq:PA2a}) and (\ref{eq:PA2b}). 
While the latter demonstrates a behavior similar to that observed in the small bubble dynamics 
[cf. Eq. (\ref{eq:PA1})], the former predicts a different scaling behavior.

Results from numerical simulations are displayed in Figs. \ref{Fig3} - \ref{Fig4}. 
For a fixed value of $C_2<0$,  on increasing $C_1$, Fig. \ref{Fig3} shows the narrowing of  $P(A|x_0)$
and the displacement of its center towards smaller $A$ values, thereby reflecting the increased importance of 
the entropy term in the free energy function.
On fixing $C_1$ and on increasing $C_2$ from negative to (small) positive values, the area 
pdf develops an increasing contribution at large $A$ values, thereby hinting at the onset of large bubbles and long first-passage times when $T>T_m$.


\subsection{Distribution of the maximum before the first-passage time: $P(M)$}
\label{M-large}

The procedure here proceeds as in Section \ref{M-small}. One first finds the
cumulative probability $q(x_0)$,  
which satisfies \cite{satya-tf,Majumdar-rev2}
\begin{equation}
\frac{1}{2}\frac{d^2q(x_0)}{dx_0^2}+C_2\frac{dq(x_0)}{dx_0}=0.
\end{equation}
With the boundary conditions
 (i) $q(0)=1$ and (ii) $q(M)=0$, this gives the solution
\begin{equation}
q(x_0)=\frac{e^{-C_2x_0}\sinh[C_2(M-x_0)]}{\sinh (C_2M)}.
\label{eq:qbig}
\end{equation}
The desired pdf is obtained as the derivative of $q(x_0)$ with respect to $M$,
\begin{equation}
P(M)=\frac{C_2e^{-C_2x_0}\sinh(C_2x_0)}{\sinh^2(C_2M)};~~~~~~~~M \ge x_0.
\label{eq:PMlarge}
\end{equation}
This result has been obtained in \cite{satya-tf,Majumdar-rev2} in a different context.
Figure \ref{Fig3} compares this form with numerical results
at temperatures below $T_m$  ($C_2 <0$). 
We observe a good agreement for initial bubble sizes satisfying $x_0 > x_{\mathrm{ch}}$.
Figure \ref{Fig4} further displays numerical results, both below $T_m$ and above it, by adopting the full bubble potential.

\subsection{Joint pdf of the maximum $M$ and the corresponding time $t_m$ before the first-passage time: $P(M,t_m)$}
\label{tmM-large}

This distribution has been studied in \cite{Majumdar-rev2} for the random motion (\ref{eq:langlarge}) with $C_2 <0$
by employing the PD method described in Section \ref{PDmethod}.
To compute $P(M,t_m)$ for general $C_2$, we follow the discussion of Section \ref{tmM-small} 
and split a representative path into two parts, before and after $t_m$. 
We compute the weight of each part separately.
The weight $W_R$ of the path after $t_m$ is given by $W_R=q(M-\epsilon)$.
Using Eq. (\ref{eq:qbig}), and then taking $\epsilon$ to be infinitesimally
small, we get
\begin{equation}
W_R=\frac{e^{-C_2(M-\epsilon)}\sinh(C_2\epsilon)}{\sinh(C_2 M)}=\epsilon \frac{C_2e^{-C_2 M}}{\sinh (C_2M)}+{\cal O}(\epsilon^2).
\label{eq:WRlarge}
\end{equation}
The weight $W_L$ of the part before $t_m$ is obtained from Eq. (\ref{eq:WL}) by substituting $C_1=0$,
\begin{equation}
W_L \propto e^{C_2(M-\epsilon-x_0)}\sum_{p=1}^\infty e^{-E_pt_m}\psi_p(M-\epsilon)\psi_p(x_0).
\label{eq:WLlarge}
\end{equation}
Here, $\psi_p(x)$ and $E_p(x)$ are the eigenfunctions and energies of the Hamiltonian (\ref{eq:H}) 
with the potential, 
\begin{equation}
V(x)=\left\{ \begin{array}{ll}
          \frac{C_2^2}{2} & \mbox{if~} 0 < x < M, \\
          \infty & \mbox{if~} x=0,  x=M.
          \end{array}
          \right.
\label{eq:Vlarge}          
\end{equation}
The normalized eigenfunctions are easily obtained as $\psi_p(x)=\sqrt{2/M}\sin(p\pi x/M)$ 
with the corresponding energies $E_p=C_2^2/2+p^2\pi^2/(2M^2)$. 
Substituting this in Eq. (\ref{eq:WLlarge}), we get 
\begin{eqnarray}
&&W_L \propto \epsilon e^{C_2(M-x_0)-C_2^2t_m/2}\frac{2\pi}{M^2} \nonumber \\
&&\times \sum_{p=1}^\infty(-1)^{p+1}p\sin\Big(\frac{p\pi x_0}{M}\Big)e^{-p^2\pi^2t_m/(2M^2)}+{\cal O}(\epsilon^2). \nonumber \\
\end{eqnarray}
The last equation, together with Eq. (\ref{eq:WRlarge}), gives the total
probability,
\begin{eqnarray}
P(M,t_m; \epsilon)=B(\epsilon) W_L  W_R.
\label{eq:PMtmepslarge}
\end{eqnarray}
The normalization constant $B(\epsilon)$ is determined by requiring  that
$\lim_{\epsilon \rightarrow 0} \int_0^\infty P(M,t_m; \epsilon)
dt_m\rightarrow P(M)$, given in Eq. (\ref{eq:PMlarge}). 
With $\sum_{p=1}^\infty(-1)^{p-1}p\sin (px)/(p^2+a^2)=(\pi/2)\sinh (ax)/\sinh (a\pi)$ 
for $-\pi < a < \pi$ \cite{gradshteyn}, 
one gets $B(\epsilon)=1/2\epsilon^2$. Using this expression for $B(\epsilon)$
in Eq. (\ref{eq:PMtmepslarge}) and taking the limit $\epsilon \rightarrow 0$,
we get the desired pdf,
\begin{eqnarray}
&&P(M,t_m) = \frac{C_2 e^{-C_2x_0-C_2^2t_m/2}}{\sinh (C_2M)}\frac{\pi}{M^2} \nonumber \\
&&\times \sum_{p=1}^\infty(-1)^{p+1}p\sin\left(\frac{p\pi x_0}{M}\right)e^{-p^2\pi^2t_m/(2M^2)}.
\label{eq:PMtmlarge}
\end{eqnarray}
The marginal distribution $P(t_m)$ can be obtained by integrating $P(M,t_m)$ over $M$ from $x_0$ to infinity. 
The large-$t_m$ and small-$t_m$ asymptotic behaviors of $P(t_m)$ for $C_2 <0$ are discussed in \cite{Majumdar-rev2}.


\begin{widetext}
\vspace{6mm} 
\begin{tabular}{|l|l|l|}
\hline
& & \\
\multirow{1}{*}{Quantities} 
& \,\,\,\,\,\,\,\,\, Results for small bubble & \,\,\,\,\,\,\,\,\ Results for large bubble \\ 
& & \\
\hline
\hline
\multirow{6}{*}{$P(t_f|x_0) $}
& & \\
& $ \sim t_f^{-C_1-3/2}e^{-x_0^2/2t_f}$ &  $\sim t_f^{-3/2}e^{-(x_0+C_2t_f)^2/2t_f}$   \\
&  $\sim t_f^{-C_1-3/2}$ ; \,\,\,\, $t_f\rightarrow \infty $    &  $\sim t_f^{-3/2} e^{-C_2 x_0} e^{-C_2^2 t_f/2}$ ; \,\,\,\,
$t_f \rightarrow \infty $  \\
& $\sim e^{-x_0^2/2t_f}$ ; \,\,\,\, $t_f\rightarrow 0$  &  $\sim e^{-(x_0+C_2t_f)^2/2t_f}$ ; \,\,\,\, $t_f\rightarrow 0$  \\
& & \\
 \hline
\multirow{6}{*}{$C(x_0,t) $}
& & \\
& $ \sim t^{-C_1-1/2}$; \,\,\,\, $t\rightarrow \infty\,\,\,$
& $\sim t^{-3/2} e^{-\left[\frac{(x_0-|C_2|t)^{2}}{2t}\right]}$; \,\,\,\, $t\rightarrow \infty$ \\
& & \\
& $ \approx 1-\frac{(x_0^2/2)^{C_1-1/2}}{\Gamma(C_1+1/2)}t^{1/2-C_1} e^{-x_0^2/2t}$;\,\,\,\, $t\rightarrow 0$ & \\
& & \\
\hline
\multirow{6}{*}{$P(A|x_0) $}
& & \\
& $\sim A^{-\frac{2}{3}(C_1+2)}e^{-2x_0^3/9A}$ &  \\
&  $\sim A^{-\frac{2}{3}(C_1+2)}$ ; \,\,\,\, $A\rightarrow \infty$    &  $A^{-3/4}e^{-\left(\frac{8}{3}\right)^{1/2}|C_2|^{3/2}A^{1/2}}$; \,\,\,\, $A\rightarrow \infty$  \\
& $\sim e^{-2x_0^3/9A}$ ; \,\,\,\, $A\rightarrow 0$  &  $\sim A ^{-4/3}e^{-2x_0^3/9A}$ ; \,\,\,\, $A\rightarrow 0$  \\
& & \\
\hline
\multirow{3}{*}{$P(M) $} 
& & \\
& $\sim M^{-2C_1-2}$ & $\sim \frac{1}{\sinh^2(C_2M)}$ \\ 
& & \\
\hline
\multirow{3}{*}{$P(M,t_m) $} 
& & \\
& $\sim M^{-(C+3)} \sum\limits_{p=1}^\infty u_{\alpha p}\frac{e^{-u_{\alpha,p}^2
t_m/2M^2}}{J_{\alpha+1}(u_{\alpha p})} J_{\alpha}\left
(\frac{u_{\alpha,p}x_0}{M}\right)$, 
&  $\sim \frac{e^{-C_2^2t_m/2}}{M^2 \sinh(C_2M)} \sum\limits_{p=1}^\infty (-1)^{p+1} p \sin\left ( \frac{p \pi
x_0}{M}\right )e^{-p^2\pi^2t_m/2M^2}$ \\
& & \\
& where $C=2C_1 +1 -\sqrt{1+4C_1^2}/2$ & \\
& & \\
\hline 
\end{tabular}
\vspace{3mm} 
\newline  \hspace{5mm}  Table 1: Scaling behavior of the probability
distribution functions of various Brownian functionals calculated in
this work for small and large DNA bubbles.
\vspace{5mm} 
\end{widetext}

\section{Conclusions}
\label{conclusions}
In this paper, we derived probability distribution functions of various Brownian 
functionals associated with a random walk model for DNA bubble dynamics at temperatures below, 
at, and above the denaturation temperature. 
Based on the backward Fokker-Planck method discussed in \cite{Majumdar-rev}, 
we derived (i) the first-passage time distribution $P(t_f|x_0)$, providing information about the bubble lifetime, 
(ii) the distribution $P(A|x_0)$, of the area $A$ covered by the random 
walk till the first-passage time, measuring the bubble reactivity to processes within the DNA, 
and (iii) the distribution $P(M)$, of the maximum bubble size $M$ before bubble closure, 
all conditioned on an initial bubble of size $x_0$ ($x_0 \in [0,M]$).
(iv) The joint probability distribution $P(M,t_m)$ of the maximum bubble size $M$ and the time $t_m$ of its occurrence before the 
first passage time was also obtained by employing the Feynman-Kac path integral formulation. 
The advantage of the elegant methods adopted here is that they produce results on various 
functionals by making proper choices of a single term in a parent differential equation with 
appropriate boundary conditions.

We considered separately the dynamics of small and large bubbles.  
Analytical results for the pdfs at each limit nicely  agree with Langevin simulations.
Our analysis reveals different nontrivial scaling behaviors of 
$P(t_f|x_0)$, $P(A|x_0)$, $P(M)$ and $P(M,t_m)$, as summarized in Table 1.
The scaling exponents are characterized either by the entropic parameter $C_1$, 
or by the base-pair dissociation parameter $C_2$. 
These quantities may thus be estimated experimentally by using fluorescence correlation spectroscopy \cite{Oleg} 
to measure,  e.g., the maximum size distribution $P(M)$ for small and large bubbles separately.

We expect our results to be useful in quantifying chemical processes within DNA, 
for example, protein binding to single-stranded DNA, and for developing a deeper understanding of polymer dynamics. 
It is of interest to extend our study and consider loop-loop interactions \cite{c-refs}, and 
the effects of disorder and heterogeneity in predicting the kinetics of specific processes within DNA bubbles 
\cite{dis0,hetero0,hetero1,hetero2}.


\begin{acknowledgments}
DS and MB acknowledge support from the Connaught fund and from NSERC. 
SG thanks Amir Bar, Rapha\"{e}l Chetrite, Ori Hirschberg, Satya N. Majumdar and David Mukamel for fruitful discussions and suggestions, 
and gratefully acknowledges the Israel Science Foundation (ISF) for supporting his research at the Weizmann Institute of Science. 
\end{acknowledgments}



\begin{thebibliography}{99}

\bibitem{DNA1}
J. D. Watson and F. H. C. Crick, Cold Spring Harb. Symp. Quant. Biol. {\bf 18}, 123 (1953).

\bibitem{DNA2}
A. Kornberg and T. A. Baker, {\em DNA Replication Second Edition} (University Science Books, Sausalito, California, 2005).

\bibitem{DNA3}
M. D. Frank-Kamenetskii, Phys. Rep. {\bf 288}, 13 (1997).

\bibitem{DNA-heat}
R. M. Wartell and A. S. Benight, Phys. Rep. {\bf 126}, 67 (1985).

\bibitem{Poland}
D. Poland and H. A. Scheraga, {\em Theory of Helix-Coil Transitions in Bio-Polymers} (Academic Press, New York, 1970).

\bibitem{DNA-force}
C. Danilowicz, Y. Kafri, R. S. Conroy, V. W. Coljee, J. Weeks, and M. Prentiss, Phys. Rev. Lett. {\bf 93}, 078101 (2004).

\bibitem{Oleg-rev}
O. Krichevsky and G. Bonnet, Rep. Prog. Phys. {\bf 65}, 251 (2002).

\bibitem{Oleg}
G. Altan-Bonnet, A. Libchaber, and O. Krichevsky, Phys. Rev. Lett.
{\bf 90}, 138101 (2003).

\bibitem{Special}
J. Phys.: Condens. Matter, Special Section on DNA melting {\bf 21} (2009).

\bibitem{Metzler-rev}
R. Metzler, T. Ambj\"ornsson, A. Hanke, and H. C. Fogedby, J. Phys.: Condens. Matter {\bf 21}, 034111 (2009), and references therein.

\bibitem{master1}
H. Kunz, R. Livi, and A. S\"{u}to, J. Stat. Mech.: Theory Exp. P06004 (2007).

\bibitem{master2}
T. Ambj\"{o}rnsson, S. K. Banik, M. A. Lomholt, and R. Metzler, Phys. Rev. E {\bf 75}, 021908 (2007).

\bibitem{Banik}
S. K. Banik, T. Ambj\"{o}rnsson, and R. Metzler, Europhys. Lett. {\bf 71}, 852 (2005).

\bibitem{FP1}
A. Hanke, and R. Metzler, J. Phys. A: Math. Gen. {\bf 36}, L473 (2003). 

\bibitem{FP2}
A. Bar, Y. Kafri, and D. Mukamel, Phys. Rev. Lett. {\bf 98}, 038103 (2007).

\bibitem{FP3}
A. Bar, Y. Kafri, and D. Mukamel, J. Phys.: Condens. Matter {\bf 21}, 034110 (2009).

\bibitem{Bishop}
T. Dauxois, M. Peyrard, and A. R. Bishop, Phys. Rev. E {\bf 47}, 684 (1993).

\bibitem{FP4}
S. Srivastava and Y. Singh, Europhys. Lett. {\bf 85}, 38001 (2009).

\bibitem{Metzler-col}
H. C. Fogedby and R. Metzler, Phys. Rev. Lett. {\bf 98}, 070601 (2007); 
Phys. Rev. E {\bf 76}, 061915 (2007).

\bibitem{Wu}
L.-A. Wu, S. S. Wu, and D. Segal, Phys. Rev. E {\bf 79}, 061901 (2009).


\bibitem{commfunc}
Note that we refer to the functionals here as ``Brownian functionals", although the dynamics 
involves both the random delta-correlated force 
and a deterministic component 
corresponding to the bubble potential, see Eq. (\ref{eq:free}).

\bibitem{Majumdar-rev}
S. N. Majumdar, Current Science {\bf 89}, 2076 (2005).

\bibitem{Majumdar-rev2}
J. Randon-Furling and S. N. Majumdar, J. Stat. Mech.: Theory Exp. P10008 (2007).

\bibitem{Kac}
M. Kac, Trans. Am. Math. Soc. {\bf 65}, 1 (1949).

\bibitem{majumdar1}
S. N. Majumdar and M. J. Kearney, Phys. Rev. E {\bf 76}, 031130 (2007). 

\bibitem{majumdar2}
P. L. Krapivsky, S. N. Majumdar and A. Rosso, J. Phys. A: Math. Theor. {\bf 43}, 315001 (2010). 

\bibitem{c-refs}
Y. Kafri, D. Mukamel, and L. Politi, Physica A {\bf 306}, 39 (2002).

\bibitem{infi}
As discussed in \cite{Majumdar-rev2}, one is required to impose an infinitesimal cut-off $\epsilon$,
or else an infinite number of crossings of the level $M$ (immediately after the first contact at time $t_m$)
would make it impossible to ensure that the path stays below $M$ at times subsequent to $t_m$.


\bibitem{algo}
A. C. Bra\'{n}ka and D. M. Heyes, Phys. Rev. E {\bf 58}, 2611 (1998).

\bibitem{gradshteyn}
I. S. Gradshteyn and I. M. Ryzhik, {\em Tables of Integrals, Series, and Products} (Academic Press, New York, 1965). 

\bibitem{arfken}
G. B. Arfken and H. J. Weber, {\em Mathematical Methods for Physicists Fourth Edition} (Academic Press, London, 1995).

\bibitem{bateman}
{\em Tables of Integral Transforms, Volume I, Based, in part, on notes left by Harry Bateman and compiled by the staff of the Bateman Manuscript Project}, edited by A. Erd\'{e}lyi, M. F. Oberhettinger and F. G. Tricomi (McGraw-Hill, New York, 1954).

\bibitem{bray}
A. J. Bray, Phys. Rev. E {\bf 62}, 103 (2000). 

\bibitem{bowman}{\em Introduction to Bessel Functions}, F. Bowman (Dover, New York, 1958).


\bibitem{satya-tf}
M. J. Kearney and S. N. Majumdar, J. Phys. A: Math. Gen. {\bf 38}, 4097 (2005).


\bibitem{prudnikov}
A. P. Prudnikov, Yu. A. Brychkov and O. I. Marichev {\em Integrals and Series: Volume 2 Special Functions}
(Gordon and Breach Science Publishers, Netherlands, 1998).

\bibitem{satya-A}
M. J. Kearney, S. N. Majumdar, and R. J. Martin, J. Phys. A: Math. Theor. {\bf 40}, F863 (2007).

\bibitem{dis0}
B. Coluzzi and E. Yeramian, Eur. Phys. J. B {\bf 56}, 349 (2007).

\bibitem{hetero0}
T. Ambj\"ornsson, S. K. Banik, O. Krichevsky, and R. Metzler,
Phys. Rev. Lett. {\bf 97}, 128105 (2006).

\bibitem{hetero1}
T. Ambj\"ornsson, S. K. Banik, O. Krichevsky, and R. Metzler, Biophys. J. {\bf 92}, 2674 (2007).

\bibitem{hetero2}
M. Zoli, Phys. Rev. E {\bf 81}, 051910 (2010).


\end{thebibliography}
\end{document}